\documentclass[a4paper,fleqn,usenatbib]{mnras}
\usepackage{newtxtext,newtxmath}
\usepackage[T1]{fontenc}
\usepackage{ae,aecompl}
\usepackage{epsfig}
\usepackage{amsmath, amsfonts, epsfig, xspace}
\usepackage{natbib}
\usepackage{rotating}
\usepackage{graphicx}
\usepackage{caption}
\usepackage{longtable}
\usepackage{multicol}
\usepackage{booktabs}
\usepackage{amsmath}
\usepackage{rotating}
\usepackage{booktabs}
\usepackage{changepage}
\usepackage{scalerel}
\def\balblzs{BAL$\_$blazars}
\def\balblz{BAL$\_$blazar}
\def\mzblazar{{\it m$_{\scaleto{\rm R \rm}{3.5pt}}$$-$z matched}~}

\def\civ{C~{\sc iv}~}
\def\aj{{AJ}}%
\def\araa{{ARA\&A}}%
\def\apj{{ApJ}}%
\def\apjl{{ApJ}}%
\def\apjs{{ApJS}}%
\def\aap{{A\&A}}%
\def\aaps{{A\&AS}}%
\def\mnras{{MNRAS}}%
\def\nar{{New A Rev.}}%
\def\pasj{{PASJ}}%

\title[]{A Search for Blazar Activity in Broad-Absorption-Line Quasars}
\author[$Mishra$ et al.]{{\Large Sapna Mishra$^{1, 2}$\thanks{E-mail: sapna@aries.res.in(SM)}, 
Gopal-Krishna$^{3}$, Hum Chand$^{4}$, Krishan Chand$^{1}$, Amit Kumar$^{1}$, and Vibhore Negi$^{1}$}\\\\
  $^{1}$Aryabhatta Research Institute of Observational Sciences (ARIES), Manora Peak, Nainital $-$ 263002, India\\
  $^{2}$Department of Physics \& Astrophysics, University of Delhi, Delhi 110 007, India\\
  $^{3}$UM-DAE Centre for Excellence in Basic Sciences, Vidyanagari, Mumbai-400098, India\\
  $^{4}$Department of Physics and Astronomical Sciences, Central University of Himachal Pradesh (CUHP), Dharamshala-176215, India\\
}

\begin{document}
\date{Accepted ---. Received ---; in original form ---}

\pagerange{\pageref{firstpage}--\pageref{lastpage}} \pubyear{2021}

\maketitle

\label{firstpage}
\begin{abstract}
  
Our recently reported lack of Intra-Night Optical Variability (INOV) among Broad-Absorption-Line (BAL) quasars exhibiting some blazar-like radio properties, either questions polar ejection of BAL clouds, and/or hints at a physical state of the relativistic jet modified due to interaction with the thermal BAL wind on the innermost sub-parsec scale. As a robust check on this unexpected finding for the BAL\_blazar candidates, we report here the INOV study of a new and much more rigorously defined comparison sample consisting of 9 normal (non-BAL) blazars, matched in both magnitude and redshift to the aforementioned sample of BAL\_blazar candidates. The present campaign spanning 27 sessions yields an INOV duty cycle of $\sim$23\% for the comparison sample of normal blazars, employing the {\it enhanced F-test}. However, even this more sensitive test does not detect INOV for the sample of BAL\_blazar candidates. This stark INOV contrast found here between the BAL\_blazar candidates and normal blazars can probably be traced to a physical interaction of the relativistic jet with the thermal wind, within sub-parsec range from the nucleus. The consequent enfeebling of the jet would additionally explain the striking deficiency among BAL quasars of powerful FR II radio lobes on the much larger scale of 10$-$100 kpc.   

\end{abstract}

\section{Introduction}
\label{sec:intro_blz}
The accretion and ejection of thermal gas is a key attribute of Active Galactic Nuclei (AGN) harbouring a supermassive black hole \citep[SMBH,][]{1984ARA&A..22..471R,1993ARA&A..31..473A}. A spectacular manifestation of the gaseous outflows are the broad absorption lines (BAL) seen in the quasar spectra, with a detection rate approaching $\sim$40\% in the near-IR, but lower (15$-$20\%) in the optical/UV  \citep{2003AJ....125.1784H}. With an ejection speed of up to $\sim$ 0.1c, BAL clouds are believed to be significant contributors to the AGN `feedback' that  probably controls  the evolution of the host galaxy and even influences the regions far beyond \citep[e.g.,][]{2010MNRAS.401....7H,doi:10.1146/annurev-astro-081811-125521,2013ARA&A..51..511K,2014ARA&A..52..589H}. Relativistic jets of synchrotron plasma, ejected from many AGNs, are another major source of AGN feedback \citep[e.g.,][]{2007NewAR..51..168B,2021MNRAS.503.1780J}. Both these modes of AGN feedback, namely, the BAL wind and relativistic jets, are together observed in a tiny fraction of the most powerful AGN, namely, quasars and both originate within the innermost sub-parsec scale from the central engine \citep{2009PASJ...61.1389D,2013A&A...554A..94B,2013ApJ...772....4H}. It is therefore important to investigate their inter-dependence and any interaction between them. Early models of the BAL phenomenon did not posit a strong interaction, since the BAL wind was envisioned to be predominantly non-polar \citep{1995ApJ...451..498M,2000ApJ...545...63E,2000ApJ...538L.103V}. However, subsequent studies have indicated that, like the jets, the BAL wind can also be bi-polar, since the BAL phenomenon is also observed in the quasars viewed near the polar axis, e.g., flat-spectrum radio quasars \citep[FSRQs,][]{2006MNRAS.372L..58B,2006ApJ...639..716Z,2007ApJ...661L.139G,2008ApJ...676L..97W} and those with radio cores of high brightness temperature ( $>$10$^{12}$ K), as inferred from their radio flux variability \citep{2006ApJ...639..716Z,2007ApJ...661L.139G}. A polar thermal outflow would also be consistent with the 3-D simulations for rapidly spinning black holes \citep{2005ApJ...620..878D,2006ApJ...641..103H}. It has also been argued that both equatorial and polar outflows can even be present in a single BALQSO \citep[e.g.,][]{2006MNRAS.372L..58B,2012MNRAS.419L..74Y,2016MNRAS.456.3929S}.\par
Clearly, polar wind would have much greater relevance to the issue of interaction of the nonthermal (jet) and thermal (BAL) outflows and this could arise in multiple ways. First, of course, is the competitive partitioning of the AGN energy output between the thermal and nonthermal channels \citep{1996A&A...308..321F,1999ApJ...527..624P,2008ApJ...687..859S,2014MNRAS.445...81S}. Secondly, a physical interaction of the inner relativistic jet with the outflowing BAL clouds on sub-parsec scale can both weaken the jet via deceleration caused by thermal mass loading of the relativistic plasma jet \citep[e.g.,][and references therein]{2019MNRAS.485..872W} and, conceivably also modify its physical state, affecting the nature of internal shocks and the physics of their turbulent wakes which are the likely sites of particle acceleration and rapid intensity fluctuations \citep{2012A&A...544A..37G,2016ApJ...820...12P}, within the basic framework outlined in \citet{2008Natur.452..966M}. On the other hand, a powerful jet flow could effectively disintegrate, or even sweep aside the BAL clouds out of its way, thus curtailing the appearance of BALs \citep[e.g.,][]{2008ApJ...687..859S}.\par
Recent theoretical work posits additional factors being responsible for the geometry of the BAL wind, namely, the mass, M$_{BH}$, and Eddington accretion rate, $\dot m$ of the SMBH. As discussed by \citet{2019A&A...630A..94G}, BAL outflow is expected to be {\it quasi-isotropic} for the parameter space that is believed to be conducive for ejection of powerful relativistic jets \citep[viz., M$_{BH}$ of the order of 10$^{7}$ to 10$^{9}$ \(M_\odot\), and  $\dot m$ $\sim$ 10$^{-3}$ to 10$^{-2}$, e.g.,][]{2014ARA&A..52..529Y}. If so, one may expect to observe BAL troughs in the spectra of even that small subset of powerful quasars whose polar axis, as defined by the jets, is oriented very close to our direction. Since the close alignment would make their jet's radiation relativistically strongly beamed in our direction, we recently posed the question: {\it Are there broad absorption-line blazars?} \citep[][hereafter Paper I]{2019MNRAS.489L..42M}. In that study, we employed a strong INOV (amplitude $>$ 3 $-$ 4 \%) with a large duty cycle of $\sim$ 30 $-$ $50\%$ as a reliable proxy for blazar-type activity \citep[e.g.,][]{2018BSRSL..87..281G}. That INOV search, however, proved negative,  thus yielding no evidence for an association between the BAL and blazar phenomena (Sec.~\ref{sec:stat_ana_blz}). This, rather unexpected result hinted at several possibilities (Paper I), e.g., BAL formation could be hindered/curtailed because the inner, optically radiating part of the putative blazar jet is able to disintegrate/evaporate the gas clouds, or just sweep them aside, i.e., out of the line-of-sight (see above).  In turn, the putative cloud-evaporation could decelerate the jet via thermal mass loading, leading to apparent fading of both the optical flux and BAL troughs. In view of such potentially drastic implications of sub-parsec scale interactions for the overall jet physics, it becomes important to investigate any association existing between the BAL and blazar phenomena. Conventionally, a high optical polarization is employed as a blazar signature \citep[$p_{opt}  >$ 3\%, e.g.,][]{1980ARA&A..18..321A,1988A&A...205...86F,2000ApJ...541...66L}. However, for BALQSOs, radio polarisation may be preferred, since a fairly high optical polarization may even get imprinted in their optical/UV spectra due to scattering of the thermal optical emission from the disk by the dense BAL wind \citep{1999ApJS..125....1O,2011ApJS..193....9D}.\par
Thus, in Paper I, we reported an INOV search among 10 optically bright high-ionization \civ BAL\_blazar candidates. The sample is fairly representative of BAL\_quasar population in terms of Balnicity Index and Absorption Index. although its median value (21500 km/s) of maximum outflow speed  is near the higher end of the distribution. The sample was assembled from the literature following two primary criteria: (i) a flat or inverted radio spectrum ($\alpha > -$0.5 for f$_{\nu} \propto \nu^{\alpha}$) and (ii) a high fractional radio polarisation ($p_{rad}$ > 3\%) which places the selected objects in the high polarization tail observed for BAL quasars \citep{2018ApJ...862..151H}. We monitored each source in 3 sessions of 3$-$5 hour duration and the derived {\it differential light curves} (DLCs) were checked for INOV using the $F_{\eta}$ statistical test \citep[see, e.g.,][]{2012A&A...544A..37G}. Rather unexpectedly, INOV was not detected in any of the 30 sessions devoted to the 10 BAL\_blazar candidates and we compared this null result with the INOV statistics for a photometric-sensitivity-matched sample of 28 DLCs pertaining to 15 `normal' blazars (i.e., lacking a BAL). This comparison sample of blazars was culled from the large INOV survey  reported by \citet{2013MNRAS.435.1300G}, also using the R-filter and employing the same statistical test ($F_{\eta}$, see Paper I for details). As expected, a high INOV duty cycle with DC $\sim$ 40\%  was found for the comparison sample comprising of normal blazars, underscoring the stark contrast to the sample of 10 BAL\_blazar candidates. However, a major caveat in this comparison was the systematic difference between the redshifts of the two samples. The sample of BAL\_blazar candidates has a much higher redshift (median $z$ = 2.13), as compared to the comparison sample of normal blazars (median $z$ = 0.42). Thus, even though the two blazar samples were monitored in sessions of similar duration (3$-$5 hrs), the corresponding {\it rest-frame} durations, T$_{int}$, are very different, the median values being 1.2 hr for the BAL\_blazar candidates and 4.2 hr for the comparison sample of normal blazars. Paper I discussed the likely impact of the different T$_{int}$, in view of the indications that a longer monitoring duration can be more propitious for INOV detection  \citep[e.g.,][]{1992AJ....104...15C,2002A&A...390..431R}. Thus, it was argued in Paper I that INOV DC declines rather slowly with T$_{int}$ decreasing from   $\sim$5.5 hr to   $\sim$3.0 hr. However, the available data did not permit the extension of this check to T$_{int} \sim$ 1 $-$ 2 hr  which pertains to the sample of BAL\_blazar candidates, as mentioned above. The goal of the present study is, firstly, to remedy this mismatch in T$_{int}$ by making INOV observations of a new well-matched comparison sample of normal blazars and, secondly, to apply a more sensitive statistical test \citep[{\it 'enhanced $\eta$ test', F$_{enh}$},][]{2014AJ....148...93D} for INOV detection to  {\it each} of the two blazar samples being compared.\par  
\section{The sample}
\label{sec:sample_blz}
\begin{table*}
\begin{minipage}[10]{170mm}
\scriptsize   
\caption{Properties of our \mzblazar comparison sample of 9 normal blazars.}
\label{tab:m-z_blazars}
\begin{tabular}{ccccclllll}
\hline
Source name &  RA (J2000)  &  DEC (J2000)  &  $z_{em}$  &  R$_{mag}^{\dagger}$  &  S$_{1.4 GHz}$  &   S$_{5 GHz}$ & $\alpha_{radio}$  &  Fractional  & Reference  \\
            &  hh:mm:ss    &  $\degr$ : $\arcmin$ : $\farcs$ &          &  (ROMABZ)  &  (mJy, ROMABZ)   &  (mJy, ROMABZ) & ($f_{\nu} \propto \nu^{\alpha}$)     &  polarisation ($\%$)  &    \\
\hline
5BZQJ0231$+$1322 & 02:31:45.89 & $+$13:22:54.7 & 2.07 & 17.50& 1559.7$\pm$46.8  &  2608$\pm$232   &  0.40  &  2.10$\pm$0.30 ( 86  GHz) & \citet{Agudo2014}     \\   
5BZQJ0249$+$0619 & 02:49:18.01 & $+$06:19:51.9 & 1.88 & 18.00& 498.1$\pm$14.9   &  620$\pm$55     &  0.17  &  1.27$\pm$0.03 ( 8.4 GHz) & \citet{2007MNRAS.376..371J}\\   
5BZQJ0646$+$4451 & 06:46:32.03 & $+$44:51:16.6 & 3.39 & 18.30& 452.4$\pm$13.6   &  1220$\pm$108   &  0.78  &  4.40$\pm$0.30 ( 86  GHz) & \citet{Agudo2014}      \\   
5BZQJ0750$+$4814 & 07:50:20.44 & $+$48:14:53.6 & 1.96 & 18.40& 715.7$\pm$21.5   &  902$\pm$80     &  0.18  &  2.15$\pm$0.03 ( 8.4 GHz) & \citet{2007MNRAS.376..371J}\\   
5BZQJ1017$+$6116 & 10:17:25.88 & $+$61:16:27.5 & 2.81 & 18.10& 404.4$\pm$12.1   &  596$\pm$53     &  0.70  &  2.11$\pm$0.04 ( 8.4 GHz) & \citet{2007MNRAS.376..371J}\\   
5BZQJ1035$+$3756 & 10:35:51.17 & $+$37:56:41.7 & 1.51 & 17.00& 52.5$\pm$1.6     &  34$\pm$5       &  $-$0.34  &  0.82$\pm$0.56 ( 8.4 GHz) & \citet{2007MNRAS.376..371J}\\   
5BZQJ1125$+$2610 & 11:25:53.70 & $+$26:10:19.9 & 2.34 & 18.20& 921.2$\pm$27.6   &  1176$\pm$104   &  0.19     &  \multicolumn{1}{l}{12.0}( 5 GHz) & \citet{2007ApJ...658..203H}\\ 
5BZQJ1126$+$4516 & 11:26:57.65 & $+$45:16:06.3 & 1.81 & 17.20&404.0$\pm$12.1   &  360$\pm$32     &  $-$0.09  &  1.32$\pm$0.04 ( 8.4 GHz) & \citet{2007MNRAS.376..371J}\\   
5BZQJ1306$+$4741 & 13:06:29.94 & $+$47:41:32.4 & 2.52 & 18.30&53.6$\pm$1.7     &  56$\pm$6       &  0.02     &  -            & -              \\   
\hline
\multicolumn{9}{l}{\scriptsize $^{\dagger}$ R$_{mag}$ in ROMABZ is from USNO-B1 catalog \citep{2003AJ....125..984M} where photometric accuracy is typically 0.25 mag.}\\

\end{tabular}
\end{minipage}
\end{table*}
The new comparison sample for the set of 10 BAL\_blazar candidates reported in Paper I, is drawn out of the blazar catalog published by \citet[][hereafter ROMA-BZ]{2009A&A...495..691M}. This catalog comprises 3561 objects. Limiting to just those classified as `RL\_FSRQ' or 'BLLac', resulted in a set of 2487 sources. By further limiting to those with a positive declination and m$_{\scaleto{\rm R \rm}{3.5pt}}$ $<$ 18.5 mag, as suited for intra-night monitoring with the 1-meter class telescopes of ARIES, we were left with 885 blazars. In this list, we searched for one normal (i.e., lacking a BAL) blazar counterpart to each of the 10 BAL\_blazar candidates, by matching within narrow windows of redshift ($\Delta \rm z = \pm$0.1) and R-magnitude ($\Delta \rm m = \pm$0.5 mag). A matching normal blazar could thus be found for each BAL\_blazar candidate, with the sole exception of J1054$+$5123. This m$_{\scaleto{\rm R \rm}{3.5pt}}-$z matched sample of 9 normal blazars (Table~\ref{tab:m-z_blazars}) forms our new and robustly defined comparison sample for the 10 BAL\_blazar candidates. In addition, all 19 blazars in our two samples are undetected in the fourth Fermi Large Area Telescope catalog \citep{2020ApJS..247...33A} and are unresolved at 1.4 GHz on a few arcsecond scale, based on their peak and integrated flux densities in the FIRST survey. Thus in radio compactness \citep{2008AJ....136..684K}, the two samples are statistically indistinguishable. Here we present results of our intra-night monitoring of all 9 normal blazars constituting the new comparison sample. In addition, our entire INOV data, both for the present comparison sample, as well as the set of 10 BAL\_blazar candidates reported in Paper I, are interpreted here by applying the more powerful statistical test, the {\it 'enhanced F-test'}\citep{2014AJ....148...93D}.
\section{Photometric monitoring}
\label{sec:obs_blz}
R-band photometric monitoring of the comparison sample of 9 normal blazars was carried out in 27 sessions (i.e., 3 sessions per source), each lasting > 3 hours. The 1.3 meter Devasthal Fast Optical Telescope \citep[DFOT,][]{2011CSci..101.1020S}, operated by the Aryabhatta Research Institute of observational sciencES (ARIES), Nainital (India), was used for 22 of the sessions. The telescope has the Ritchey-Chretien (RC) optics with f/4 Cassegrain focus, yielding a plate scale of 40 arcsec-mm$^{-1}$ and a pointing accuracy of < 10 arcsec (rms). The data were recorded on a Peltier-cooled Andor CCD camera having 2048 $\times$ 2048 pixels of 13.5 $\mu$m size, covering an 18$\times$18 arcmin$^{2}$ field-of-view (FoV). The CCD detector has a gain of 2 e$^{-}$ per analog-to-digital unit (ADU) and a readout noise of 7e $^{-}$ at a speed of 1000 kHz. The camera is cooled thermoelectrically down to $-$85$^{o}$C. For another 4 of the sessions, the 2.0-meter Himalayan Chandra Telescope (HCT) of the Indian Astronomical Observatory at Hanle in Ladakh (India) \citep{2010ASInC...1..193P} was used. The telescope has an RC optics with f/9 and a Cassegrain focus. It is equipped with a 2148 $\times$ 2048 cryogenically cooled detector covering a 10 $\times$ 10 arcmin$^{2}$ FoV. The plate scale of the detector is 0.296 arcsec per pixel. The CCD has readout noise of 5.75 e$^{-}$ per pixel, and a gain of 0.28 e$^{-}$/ADU. Monitoring during the remaining one session was carried out with the 1.04 meter Sampurnanand Telescope \citep[ST,][]{1999CSci...77..643S}, operated by ARIES. The telescope is an equitorially mounted, f/13 RC reflector with a Cassegrain focus. The data were recorded on a 4K $\times$ 4K Imager of 15 $\mu$m pixel size. The detector covers a 15 $\times$ 15 arcmin$^{2}$ area of the sky. The readout noise and gain of this detector are 10 e$^{-}$ and 3 e$^{-}$/ADU, respectively. In each session, one normal blazar of the comparison sample was monitored continuously for minimum 3 hours, with a typical exposure of 5-10 min per frame. Typically, the seeing value during our monitoring sessions was $\sim$ 3.5 arcsec (online Figures 1, 2, and 3). Details of the comparison stars used for deriving the DLCs of the 9 normal blazars are provided in the online Table 1. Preliminary image processing (bias subtraction, flat-fielding, and cosmic-ray removal) and aperture photometry for each observing session was performed using similar procedure described in Paper I.
\section{Statistical analysis and Result}
\label{sec:stat_ana_blz}
To ascertain the presence of INOV in our \balblz~ and comparison samples, we subjected their DLCs (relative to the three comparison stars, in each session) to the {\it `enhanced F-test'} \citep{2014AJ....148...93D,2016MNRAS.460.3950P}. The main advantage of this test is that it modifies the DLCs of the comparison stars in such a way that they have the same photometric noise, making their magnitudes appear the same as the mean magnitude of the target AGN/quasar, thus ensuring that the analysis is not affected by any magnitude difference between the AGN and the comparison star(s). Furthermore, the power of this test is enhanced since more than two comparison stars are generally used, the closest one in magnitude to the quasar being taken as the reference star. The statistical criterion for the {\it `enhanced F-test'} is defined as:
\begin{equation}
  \label{eq.ftest3}
  F_{enh} = \frac{Var(q-ref)}{Var_c},~~ Var_c=\frac{1}{\sum_{j=1}^k N_j - k}\sum_{j=1}^{k}\sum_{i=1}^{N_j}s_{j,i}^{2}
\end{equation}
where, $Var(q-ref)$ is the variance of the {\it quasar-reference star} DLC and $Var_c$ is the stacked variance \citep{2014AJ....148...93D} of the {\it comparison star $-$ reference star} DLCs, $N_j$ is the number of observations of the $j^{th}$ comparison star, $k$ is the total number of comparison stars. $s_{j,i}^2$ is the scaled square deviation, defined as
\begin{equation}
  s_{j,i}^{2}=\omega_j(m_{j,i}-\bar{m_j})^{2},~~\\
  \omega_j=\frac{<\sigma_{err}^{2}(q-ref)>}{<\sigma_{err}^{2}(s_j-ref)>} \nonumber
\end{equation}
Here, $m_{j,i}$'s and $\bar{m_j}$ are the differential instrumental magnitudes and the mean differential magnitude of the $j^{th}$ comparison star$-$reference star DLC.
And  <$\sigma_{err}^2(q-ref)$>  and  <$\sigma_{err}^2(s_j-ref)$> are, respectively, the session-averaged variances of the `quasar$-$reference star' and the `$j^{th}$comparison$-$reference star' DLCs, based on the rms photometric errors ($\sigma_{err}$) returned by the DAOPHOT for individual photometric measurements.\par
\begin{figure}
\includegraphics[width=0.48\textwidth,height=0.22\textheight]{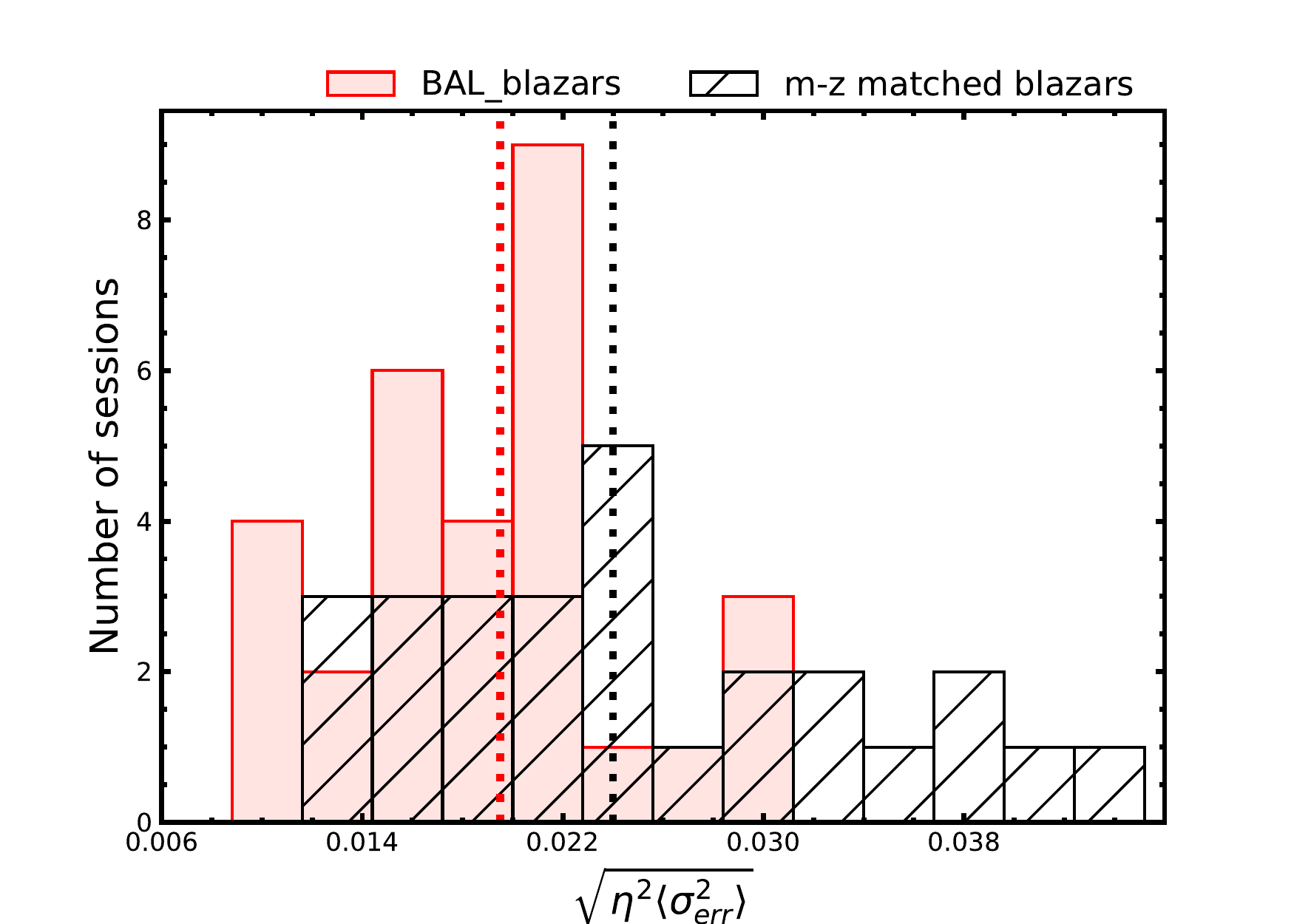}
\caption{Distributions of photometric noise parameter (PNP = $\sqrt { \eta^2 \langle \sigma^2_{err} \rangle}$) for the 30 monitoring sessions devoted to the 10 \balblzs~of Paper I (red shaded) and the 27 sessions devoted to the matched comparison sample of 9  normal blazars monitored in the present study (black). The two dotted vertical lines mark the median values for the respective histograms.}
\label{fig:rms_comp_blz}
\end{figure}

For each DLC, online Table 2 and Table 3 provide a comparison of our computed values of $F_{enh}$ with the critical value of $F$ ($F_{c}^{\alpha}$)  for $\alpha = $ 0.01, which corresponds to 99\% confidence level for INOV detection (online Tables 2 and 3, columns 7 and 8). If the computed value of F$_{enh}$ for a DLC exceeds the critical value, the null hypothesis (i.e., no variability) is discarded. We thus classify a DLC with $F_{enh}$ $\ge$ $F_{c}$(0.99) as variable (`V') at a confidence level $\ge$ 99\%. Column 10 in the online Tables 2 and 3 list the `Photometric Noise Parameter'' (hereafter PNP) = $\sqrt {\eta^2 \langle \sigma^2_{i,err} \rangle}$, where $\eta$ = 1.5 \citep[cf.][]{2013MNRAS.435.1300G}. Note that PNP has been averaged over the three DLCs of the target blazar in that session; it is basically the inverse of the photometric sensitivity attained in that session (Fig.~\ref{fig:rms_comp_blz}).

\subsection{Estimation of the INOV duty cycle}
\label{sec:DC_estimation_blz}

As in Paper I, we have used the definition of \citet{1999A&AS..135..477R} for computing the duty cycle (DC) of INOV, which is given by:

\begin{equation}
\hspace{0.8in}    
DC = 100\frac{\sum_\mathbf{i=1}^\mathbf{n}
  K^i(1/T_{int}^i)}{\sum_\mathbf{i=1}^\mathbf{n}(1/T_{int}^i)} {\rm
  per cent}
\label{eq:dc} 
\end{equation}

where T$_{int}^i = T_{obs}^{i}(1+z_{em})^{-1}$ is the intrinsic duration of the $i^{th}$ monitoring session of duration T$_{obs}^{i}$. Since the Doppler factor values for our blazars from both the samples are currently unknown, their impact on the intrinsic duration is not considered in the present study. $K^{i}$ was taken as unity in case of positive INOV detection for the $i^{th}$ session, otherwise $K^{i}$ was set to zero. We find the INOV DC for the comparison sample of 9 normal blazars to be 22.7\%. At the same time, applying this {\it `enhanced $F$-test'} to the DLCs of the 10 \balblz~ (Paper I) still yields a DC of zero, unchanged from that reported in Paper I based on the (less sensitive) $F_\eta$ test.\par
\section{Discussion and Conclusions}
\label{sec:discussion_blz}

To recapitulate, the main objective of this study, in conjunction with Paper I, is to obtain the observational clues about the physical interaction which is expected to occur between the two principal modes of kinetic energy injection by AGN. Such an opportunity is presented by radio-loud broad-absorption-line QSOs (RL\_BALQSOs) since both thermal and relativistic feedback modes are strongly at work, simultaneously. Our focus here is on their tiny subset in which the nonthermal relativistic jets are expected to be oriented close to the line-of-sight, hence blazar-like, providing a direct (nearly pole-on) view of the putative interaction of the jet with the thermal wind which manifests itself as BAL troughs in the spectra. Recently, in Paper I we reported a search for blazar-like jets in a well-defined sample of 10 `BAL\_blazar' candidates, employing strong INOV as a reliable signature of a blazar jet \citep[e.g.,][and references therein]{2018BSRSL..87..281G}. That sample was extracted from  RL\_BALQSOs,  chiefly on the criterion of a high (radio) polarisation; optical polarisation data being not only even more scarce, but also because a high optical polarisation may well arise from scattering of the thermal nuclear optical emission by the dense BAL wind (Sec.~\ref{sec:intro_blz}). The unexpected finding from our INOV study of the 10  BAL\_blazar candidates was the non-detection of INOV in even one of the 30 monitoring sessions (Paper I). This placed a question mark on the existence of BAL\_blazars, with potential implications for the models invoking  polar ejection of BAL clouds, and/or the nature of their relativistic jet flow itself. In Paper I this negative result was also contrasted with the high INOV duty cycle ($\sim$ 40\%) estimated for an unbiased comparison sample of 15 `normal' (i.e., non-BAL) blazars drawn from the INOV literature. This strong contrast hinted at mutual spatial avoidance of relativistic optical synchrotron jet and BAL clouds, inspite of both phenomena believed to be occurring on the innermost sub-parsec scale from the central engine (Sec.~\ref{sec:intro_blz}).\par

However, an important caveat in that comparison was the large redshift mis-match between the monitored 10 BAL\_blazar candidates (median $z$ = 2.13), and the comparison sample of 15 normal blazars (median $z$ = 0.42), extracted from the literature. Not only did this make the comparison sample  much less luminous intrinsically, but also the rest-frame (i.e., `intrinsic' ) durations of the monitoring sessions of the BAL\_blazar candidates were much shorter ($\sim$1.2 hr), than those for the comparison sample (median $\sim$4.2 hr). This large disparity could potentially explain the observed extreme rarity, in fact the total lack, of INOV detection for the BAL\_blazar candidates. In order to remedy this shortcoming, we have presented in this work INOV observations of a new, more robustly defined comparison sample of normal blazars which is matched in the $m_R - z$  plane to our sample of 10 BAL\_blazar candidates whose INOV observations were reported in Paper I. Moreover, for both samples, INOV search has now been made by employing the {\it `enhanced F-test'} \citep{2014AJ....148...93D} which is more sensitive than the $F_\eta$ test applied in Paper I. The key result found here is that the total lack of INOV detection among the BAL\_blazar candidates, reported in Paper I, persists, whereas a high INOV duty cycle of $\sim$23\% is found for the new $m_R - z$ matched comparison sample of normal (i.e., non-BAL) blazars. Note that, unlike Paper I, the redshift matching of the two samples being compared here, has ensured an excellent match between their rest-frame monitoring durations. One remaining mismatch between the two samples is that, in contrast to the sample of BAL\_blazar candidates, the comparison sample consisting of normal blazars is marked by a weaker (radio) polarization (Table~\ref{tab:m-z_blazars}), although the flat/inverted radio spectra of all the sources in the two samples do indicate a relativistically beamed radio jet (Table~\ref{tab:m-z_blazars}). In any case, any possible correction for this polarisation mismatch would probably further amplify the already stark contrast found here between the INOV occurrence in the two samples. This is because a stronger INOV is known to correlate positively with polarization \citep[][]{2004JApA...25....1S,2012A&A...544A..37G,2018BSRSL..87..281G}.\par

Lastly, a yet another observational bias which could have led to the apparent large contrast between INOV DC of the two samples, stems from the possibility of systematically lower photometric sensitivity attained for the sample of the BAL\_blazar candidates (Paper I). To check for this, we display in Fig.~\ref{fig:rms_comp_blz} histograms of the photometric noise parameter ($\sqrt { \eta^2 \langle \sigma^2_{err} \rangle}$), for the BAL\_blazar candidates (10 objects, 30 sessions) and  for the comparison sample of normal blazars (9 objects, 27 sessions).  As a rule, we list the value of this parameter for each session in the last column of the table showing the status of INOV detection for that session (see online Table 2 and Table 3). Comparison of the two histograms (e.g., see Fig.~\ref{fig:rms_comp_blz}) provides no evidence for a systematically higher sensitivity attained for the comparison sample;  if at all, the opposite may be true. Thus, on all these counts, the striking deficit of INOV (i.e., blazar-like jets) among the BAL\_blazar candidates appears to be a genuine effect.\par

To sum up,  it may be recalled that for many years BALQSOs had remained undetected at radio-frequencies \citep{1992ApJ...396..487S} and this could be understood in terms of stifling of the jet via interaction with the outflowing dense thermal plasma \citep{2006ApJ...641..210G}. However, the situation changed when weak radio counterparts were found for many BALQSOs in the FIRST survey at 1.4 GHz \citep{1997AJ....113..144B,2000ApJ...538...72B}. Still, the case for the jets' weakening via interaction with the BAL clouds persisted, manifestating on the scale of tens or hundreds of kiloparsecs, in the form of a marked deficiency of FR II type radio lobes which are generally engendered by powerful jets \citep{2006ApJ...641..210G,2008ApJ...687..859S,2014MNRAS.440.2474W}. This posed the question: are there any other observable signatures of the putative jet-cloud interaction in radio-loud BALQSOs, specially on the innermost sub-parsec scale where such a physical interactions should indeed be occurring? The results presented here address this issue and suggest that the striking lack of INOV in BAL\_blazar candidates could well be one such observational manifestation. Strong INOV, most probably arises from within the turbulent wakes of synchrotron plasma forming behind relativistic shocks in the sub-parsec scale jets of AGN \citep[e.g.,][]{2012A&A...544A..37G,2016ApJ...820...12P}, in line with the framework presented in \citet{2008Natur.452..966M}. Any thermal mass loading of the relativistic jet by the BAL outflow is expected to not only moderate the jet's speed but also weaken the relativistic shocks in it, plausibly dampening their capacity to accelerate relativistic particles to energies required for optical synchrotron radiation. Together, both these effects might be responsible for the subdued INOV of BAL\_blazar candidates, as found here. In this context, it would be particularly useful to carry out multi-epoch VLBI imaging of BAL\_blazar candidates, in search of evidence for systematically lower jet speeds and also for measuring the brightness temperatures and flux variability of their VLBI cores, towards obtaining the much needed estimates of the jets' Doppler factors \citep[e.g.,][]{2015AJ....149...46Z}. Equally, it would be desirable to employ larger telescopes to extend the INOV search to larger (likely, fainter) samples of BAL\_blazar candidates, for a more robust characterization of their INOV. Likewise, accurate measurement of radio spectra, based on quasi-simultaneous multi-frequency obsevations, would provide additional valuable input.

\vspace{-0.2in}
\section*{Acknowledgments}
We thank the anonymous referee for the constructive comments on our manuscript. G-K acknowledged a Senior Scientist fellowship from the Indian National Science Academy. The assistance from the scientific and technical staff of ARIES DFOT and ST is sincerely acknowledged. Thanks are also due to the staff of IAO (Hanle) and CREST (Hosakote), for making possible a part of the observations reported here. The facilities at IAO and CREST are operated by the Indian Institute of Astrophysics, Bengaluru. 
\vspace{-0.1in}
\section*{Data availability}
The data used in this study will be shared on reasonable request to the corresponding author.
\bibliography{references}

\begin{thebibliography}{}
\makeatletter
\relax
\def\mn@urlcharsother{\let\do\@makeother \do\$\do\&\do\#\do\^\do\_\do\%\do\~}
\def\mn@doi{\begingroup\mn@urlcharsother \@ifnextchar [ {\mn@doi@}
  {\mn@doi@[]}}
\def\mn@doi@[#1]#2{\def\@tempa{#1}\ifx\@tempa\@empty \href
  {http://dx.doi.org/#2} {doi:#2}\else \href {http://dx.doi.org/#2} {#1}\fi
  \endgroup}
\def\mn@eprint#1#2{\mn@eprint@#1:#2::\@nil}
\def\mn@eprint@arXiv#1{\href {http://arxiv.org/abs/#1} {{\tt arXiv:#1}}}
\def\mn@eprint@dblp#1{\href {http://dblp.uni-trier.de/rec/bibtex/#1.xml}
  {dblp:#1}}
\def\mn@eprint@#1:#2:#3:#4\@nil{\def\@tempa {#1}\def\@tempb {#2}\def\@tempc
  {#3}\ifx \@tempc \@empty \let \@tempc \@tempb \let \@tempb \@tempa \fi \ifx
  \@tempb \@empty \def\@tempb {arXiv}\fi \@ifundefined
  {mn@eprint@\@tempb}{\@tempb:\@tempc}{\expandafter \expandafter \csname
  mn@eprint@\@tempb\endcsname \expandafter{\@tempc}}}

\bibitem[\protect\citeauthoryear{{Abdollahi} et~al.,}{{Abdollahi}
  et~al.}{2020}]{2020ApJS..247...33A}
{Abdollahi} S.,  et~al., 2020, \mn@doi [\apjs] {10.3847/1538-4365/ab6bcb},
  \href {https://ui.adsabs.harvard.edu/abs/2020ApJS..247...33A} {247, 33}

\bibitem[\protect\citeauthoryear{{Agudo}, {Thum}, {G{\'o}mez}  \&
  {Wiesemeyer}}{{Agudo} et~al.}{2014}]{Agudo2014}
{Agudo} I.,  {Thum} C.,  {G{\'o}mez} J.~L.,   {Wiesemeyer} H.,  2014, \mn@doi
  [\aap] {10.1051/0004-6361/201423366}, \href
  {https://ui.adsabs.harvard.edu/abs/2014A&A...566A..59A} {566, A59}

\bibitem[\protect\citeauthoryear{{Angel} \& {Stockman}}{{Angel} \&
  {Stockman}}{1980}]{1980ARA&A..18..321A}
{Angel} J.~R.~P.,  {Stockman} H.~S.,  1980, \mn@doi [\araa]
  {10.1146/annurev.aa.18.090180.001541}, \href
  {https://ui.adsabs.harvard.edu/abs/1980ARA&A..18..321A} {18, 321}

\bibitem[\protect\citeauthoryear{{Antonucci}}{{Antonucci}}{1993}]{1993ARA&A..31..473A}
{Antonucci} R.,  1993, \mn@doi [\araa] {10.1146/annurev.aa.31.090193.002353},
  \href {http://adsabs.harvard.edu/abs/1993ARA%26A..31..473A} {31, 473}

\bibitem[\protect\citeauthoryear{{Barvainis} \& {Lonsdale}}{{Barvainis} \&
  {Lonsdale}}{1997}]{1997AJ....113..144B}
{Barvainis} R.,  {Lonsdale} C.,  1997, \mn@doi [\aj] {10.1086/118240}, \href
  {https://ui.adsabs.harvard.edu/abs/1997AJ....113..144B} {113, 144}

\bibitem[\protect\citeauthoryear{{Becker}, {White}, {Gregg}  \& et
  al.}{{Becker} et~al.}{2000}]{2000ApJ...538...72B}
{Becker} R.~H.,  {White} R.~L.,  {Gregg} M.~D.,   et al. 2000, \mn@doi [\apj]
  {10.1086/309099}, \href {http://adsabs.harvard.edu/abs/2000ApJ...538...72B}
  {538, 72}

\bibitem[\protect\citeauthoryear{{Best}}{{Best}}{2007}]{2007NewAR..51..168B}
{Best} P.~N.,  2007, \mn@doi [\nar] {10.1016/j.newar.2006.11.014}, \href
  {https://ui.adsabs.harvard.edu/abs/2007NewAR..51..168B} {51, 168}

\bibitem[\protect\citeauthoryear{{Brotherton}, {De Breuck}  \&
  {Schaefer}}{{Brotherton} et~al.}{2006}]{2006MNRAS.372L..58B}
{Brotherton} M.~S.,  {De Breuck} C.,   {Schaefer} J.~J.,  2006, \mn@doi
  [\mnras] {10.1111/j.1745-3933.2006.00226.x}, \href
  {http://adsabs.harvard.edu/abs/2006MNRAS.372L..58B} {372, L58}

\bibitem[\protect\citeauthoryear{{Bruni}, {Dallacasa}, {Mack},
  {Montenegro-Montes}  \& et al.}{{Bruni} et~al.}{2013}]{2013A&A...554A..94B}
{Bruni} G.,  {Dallacasa} D.,  {Mack} K.-H.,  {Montenegro-Montes} F.~M.,   et
  al. 2013, \mn@doi [\aap] {10.1051/0004-6361/201321341}, \href
  {http://adsabs.harvard.edu/abs/2013A%26A...554A..94B} {554, A94}

\bibitem[\protect\citeauthoryear{{Carini}, {Miller}, {Noble}  \& et
  al.}{{Carini} et~al.}{1992}]{1992AJ....104...15C}
{Carini} M.~T.,  {Miller} H.~R.,  {Noble} J.~C.,   et al. 1992, \mn@doi [\aj]
  {10.1086/116217}, \href {http://adsabs.harvard.edu/abs/1992AJ....104...15C}
  {104, 15}

\bibitem[\protect\citeauthoryear{{De Villiers}, {Hawley}, {Krolik}  \& et
  al.}{{De Villiers} et~al.}{2005}]{2005ApJ...620..878D}
{De Villiers} J.-P.,  {Hawley} J.~F.,  {Krolik} J.~H.,   et al. 2005, \mn@doi
  [\apj] {10.1086/427142}, \href
  {https://ui.adsabs.harvard.edu/abs/2005ApJ...620..878D} {620, 878}

\bibitem[\protect\citeauthoryear{{DiPompeo}, {Brotherton}  \& {De
  Breuck}}{{DiPompeo} et~al.}{2011}]{2011ApJS..193....9D}
{DiPompeo} M.~A.,  {Brotherton} M.~S.,   {De Breuck} C.,  2011, \mn@doi [\apjs]
  {10.1088/0067-0049/193/1/9}, \href
  {https://ui.adsabs.harvard.edu/abs/2011ApJS..193....9D} {193, 9}

\bibitem[\protect\citeauthoryear{{Doi} et~al.,}{{Doi}
  et~al.}{2009}]{2009PASJ...61.1389D}
{Doi} A.,  et~al., 2009, \mn@doi [\pasj] {10.1093/pasj/61.6.1389}, \href
  {http://adsabs.harvard.edu/abs/2009PASJ...61.1389D} {61, 1389}

\bibitem[\protect\citeauthoryear{{Elvis}}{{Elvis}}{2000}]{2000ApJ...545...63E}
{Elvis} M.,  2000, \mn@doi [\apj] {10.1086/317778}, \href
  {http://adsabs.harvard.edu/abs/2000ApJ...545...63E} {545, 63}

\bibitem[\protect\citeauthoryear{Fabian}{Fabian}{2012}]{doi:10.1146/annurev-astro-081811-125521}
Fabian A.,  2012, \mn@doi [Annual Review of Astronomy and Astrophysics]
  {10.1146/annurev-astro-081811-125521}, 50, 455

\bibitem[\protect\citeauthoryear{{Falcke} \& {Biermann}}{{Falcke} \&
  {Biermann}}{1996}]{1996A&A...308..321F}
{Falcke} H.,  {Biermann} P.~L.,  1996, \aap, \href
  {https://ui.adsabs.harvard.edu/abs/1996A&A...308..321F} {308, 321}

\bibitem[\protect\citeauthoryear{{Fugmann}}{{Fugmann}}{1988}]{1988A&A...205...86F}
{Fugmann} W.,  1988, \aap, \href
  {http://adsabs.harvard.edu/abs/1988A%26A...205...86F} {205, 86}

\bibitem[\protect\citeauthoryear{{Ghosh} \& {Punsly}}{{Ghosh} \&
  {Punsly}}{2007}]{2007ApJ...661L.139G}
{Ghosh} K.~K.,  {Punsly} B.,  2007, \mn@doi [\apjl] {10.1086/518859}, \href
  {http://adsabs.harvard.edu/abs/2007ApJ...661L.139G} {661, L139}

\bibitem[\protect\citeauthoryear{{Giustini} \& {Proga}}{{Giustini} \&
  {Proga}}{2019}]{2019A&A...630A..94G}
{Giustini} M.,  {Proga} D.,  2019, \mn@doi [\aap]
  {10.1051/0004-6361/201833810}, \href
  {https://ui.adsabs.harvard.edu/abs/2019A&A...630A..94G} {630, A94}

\bibitem[\protect\citeauthoryear{{Gopal-Krishna} \& {Wiita}}{{Gopal-Krishna} \&
  {Wiita}}{2018}]{2018BSRSL..87..281G}
{Gopal-Krishna} {Wiita} P.~J.,  2018, Bulletin de la Societe Royale des
  Sciences de Liege, \href {http://adsabs.harvard.edu/abs/2018BSRSL..87..281G}
  {87, 281}

\bibitem[\protect\citeauthoryear{{Goyal}, {Gopal-Krishna}, {Wiita}  \& et
  al.}{{Goyal} et~al.}{2012}]{2012A&A...544A..37G}
{Goyal} A.,  {Gopal-Krishna} {Wiita} P.~J.,   et al. 2012, \mn@doi [\aap]
  {10.1051/0004-6361/201218888}, \href
  {http://adsabs.harvard.edu/abs/2012A%26A...544A..37G} {544, A37}

\bibitem[\protect\citeauthoryear{{Goyal}, {Gopal-Krishna}  \& et al.}{{Goyal}
  et~al.}{2013}]{2013MNRAS.435.1300G}
{Goyal} A.,  {Gopal-Krishna} Paul~J. W.,   et al. 2013, \mn@doi [\mnras]
  {10.1093/mnras/stt1373}, \href
  {http://adsabs.harvard.edu/abs/2013MNRAS.435.1300G} {435, 1300}

\bibitem[\protect\citeauthoryear{{Gregg}, {Becker}  \& {de Vries}}{{Gregg}
  et~al.}{2006}]{2006ApJ...641..210G}
{Gregg} M.~D.,  {Becker} R.~H.,   {de Vries} W.,  2006, \mn@doi [\apj]
  {10.1086/500381}, \href {http://adsabs.harvard.edu/abs/2006ApJ...641..210G}
  {641, 210}

\bibitem[\protect\citeauthoryear{{Hawley} \& {Krolik}}{{Hawley} \&
  {Krolik}}{2006}]{2006ApJ...641..103H}
{Hawley} J.~F.,  {Krolik} J.~H.,  2006, \mn@doi [\apj] {10.1086/500385}, \href
  {https://ui.adsabs.harvard.edu/abs/2006ApJ...641..103H} {641, 103}

\bibitem[\protect\citeauthoryear{{Hayashi}, {Doi}  \& {Nagai}}{{Hayashi}
  et~al.}{2013}]{2013ApJ...772....4H}
{Hayashi} T.~J.,  {Doi} A.,   {Nagai} H.,  2013, \mn@doi [\apj]
  {10.1088/0004-637X/772/1/4}, \href
  {http://adsabs.harvard.edu/abs/2013ApJ...772....4H} {772, 4}

\bibitem[\protect\citeauthoryear{{Heckman} \& {Best}}{{Heckman} \&
  {Best}}{2014}]{2014ARA&A..52..589H}
{Heckman} T.~M.,  {Best} P.~N.,  2014, \mn@doi [\araa]
  {10.1146/annurev-astro-081913-035722}, \href
  {https://ui.adsabs.harvard.edu/abs/2014ARA&A..52..589H} {52, 589}

\bibitem[\protect\citeauthoryear{{Helmboldt} et~al.,}{{Helmboldt}
  et~al.}{2007}]{2007ApJ...658..203H}
{Helmboldt} J.~F.,  et~al., 2007, \mn@doi [\apj] {10.1086/511005}, \href
  {https://ui.adsabs.harvard.edu/abs/2007ApJ...658..203H} {658, 203}

\bibitem[\protect\citeauthoryear{{Hewett} \& {Foltz}}{{Hewett} \&
  {Foltz}}{2003}]{2003AJ....125.1784H}
{Hewett} P.~C.,  {Foltz} C.~B.,  2003, \mn@doi [\aj] {10.1086/368392}, \href
  {http://adsabs.harvard.edu/abs/2003AJ....125.1784H} {125, 1784}

\bibitem[\protect\citeauthoryear{{Hodge}, {Lister}, {Aller}  \& et al.}{{Hodge}
  et~al.}{2018}]{2018ApJ...862..151H}
{Hodge} M.~A.,  {Lister} M.~L.,  {Aller} M.~F.,   et al. 2018, \mn@doi [\apj]
  {10.3847/1538-4357/aacb2f}, \href
  {http://adsabs.harvard.edu/abs/2018ApJ...862..151H} {862, 151}

\bibitem[\protect\citeauthoryear{{Hopkins} \& {Elvis}}{{Hopkins} \&
  {Elvis}}{2010}]{2010MNRAS.401....7H}
{Hopkins} P.~F.,  {Elvis} M.,  2010, \mn@doi [\mnras]
  {10.1111/j.1365-2966.2009.15643.x}, \href
  {https://ui.adsabs.harvard.edu/abs/2010MNRAS.401....7H} {401, 7}

\bibitem[\protect\citeauthoryear{{Jackson}, {Battye}, {Browne}  \& et
  al.}{{Jackson} et~al.}{2007}]{2007MNRAS.376..371J}
{Jackson} N.,  {Battye} R.~A.,  {Browne} I.~W.~A.,   et al. 2007, \mn@doi
  [\mnras] {10.1111/j.1365-2966.2007.11442.x}, \href
  {https://ui.adsabs.harvard.edu/abs/2007MNRAS.376..371J} {376, 371}

\bibitem[\protect\citeauthoryear{{Jarvis} et~al.,}{{Jarvis}
  et~al.}{2021}]{2021MNRAS.503.1780J}
{Jarvis} M.~E.,  et~al., 2021, \mn@doi [\mnras] {10.1093/mnras/stab549}, \href
  {https://ui.adsabs.harvard.edu/abs/2021MNRAS.503.1780J} {503, 1780}

\bibitem[\protect\citeauthoryear{{Kimball} \& {Ivezi{\'c}}}{{Kimball} \&
  {Ivezi{\'c}}}{2008}]{2008AJ....136..684K}
{Kimball} A.~E.,  {Ivezi{\'c}} {\v{Z}}.,  2008, \mn@doi [\aj]
  {10.1088/0004-6256/136/2/684}, \href
  {https://ui.adsabs.harvard.edu/abs/2008AJ....136..684K} {136, 684}

\bibitem[\protect\citeauthoryear{{Kormendy} \& {Ho}}{{Kormendy} \&
  {Ho}}{2013}]{2013ARA&A..51..511K}
{Kormendy} J.,  {Ho} L.~C.,  2013, \mn@doi [\araa]
  {10.1146/annurev-astro-082708-101811}, \href
  {https://ui.adsabs.harvard.edu/abs/2013ARA&A..51..511K} {51, 511}

\bibitem[\protect\citeauthoryear{{Lister} \& {Smith}}{{Lister} \&
  {Smith}}{2000}]{2000ApJ...541...66L}
{Lister} M.~L.,  {Smith} P.~S.,  2000, \mn@doi [\apj] {10.1086/309413}, \href
  {http://adsabs.harvard.edu/abs/2000ApJ...541...66L} {541, 66}

\bibitem[\protect\citeauthoryear{{Marscher} et~al.,}{{Marscher}
  et~al.}{2008}]{2008Natur.452..966M}
{Marscher} A.~P.,  et~al., 2008, \mn@doi [\nat] {10.1038/nature06895}, \href
  {https://ui.adsabs.harvard.edu/abs/2008Natur.452..966M} {452, 966}

\bibitem[\protect\citeauthoryear{{Massaro}, {Giommi}, {Leto}  \& et
  al.}{{Massaro} et~al.}{2009}]{2009A&A...495..691M}
{Massaro} E.,  {Giommi} P.,  {Leto} C.,   et al. 2009, \mn@doi [\aap]
  {10.1051/0004-6361:200810161}, \href
  {http://adsabs.harvard.edu/abs/2009A%26A...495..691M} {495, 691}

\bibitem[\protect\citeauthoryear{{Mishra}, {Gopal-Krishna}, {Chand}  \& et
  al.}{{Mishra} et~al.}{2019}]{2019MNRAS.489L..42M}
{Mishra} S.,  {Gopal-Krishna} {Chand} H.,   et al. 2019, \mn@doi [\mnras]
  {10.1093/mnrasl/slz122}, \href
  {https://ui.adsabs.harvard.edu/abs/2019MNRAS.489L..42M} {489, L42}

\bibitem[\protect\citeauthoryear{{Monet}, {Levine}, {Canzian}  \& et
  al.}{{Monet} et~al.}{2003}]{2003AJ....125..984M}
{Monet} D.~G.,  {Levine} S.~E.,  {Canzian} B.,   et al. 2003, \mn@doi [\aj]
  {10.1086/345888}, \href
  {https://ui.adsabs.harvard.edu/abs/2003AJ....125..984M} {125, 984}

\bibitem[\protect\citeauthoryear{{Murray}, {Chiang}, {Grossman}  \&
  {Voit}}{{Murray} et~al.}{1995}]{1995ApJ...451..498M}
{Murray} N.,  {Chiang} J.,  {Grossman} S.~A.,   {Voit} G.~M.,  1995, \mn@doi
  [\apj] {10.1086/176238}, \href
  {http://adsabs.harvard.edu/abs/1995ApJ...451..498M} {451, 498}

\bibitem[\protect\citeauthoryear{{Ogle}, {Cohen}, {Miller}  \& et al.}{{Ogle}
  et~al.}{1999}]{1999ApJS..125....1O}
{Ogle} P.~M.,  {Cohen} M.~H.,  {Miller} J.~S.,   et al. 1999, \mn@doi [\apjs]
  {10.1086/313272}, \href
  {https://ui.adsabs.harvard.edu/abs/1999ApJS..125....1O} {125, 1}

\bibitem[\protect\citeauthoryear{{Polednikova} et~al.,}{{Polednikova}
  et~al.}{2016}]{2016MNRAS.460.3950P}
{Polednikova} J.,  et~al., 2016, \mn@doi [\mnras] {10.1093/mnras/stw1252},
  \href {https://ui.adsabs.harvard.edu/abs/2016MNRAS.460.3950P} {460, 3950}

\bibitem[\protect\citeauthoryear{{Pollack}, {Pauls}  \& {Wiita}}{{Pollack}
  et~al.}{2016}]{2016ApJ...820...12P}
{Pollack} M.,  {Pauls} D.,   {Wiita} P.~J.,  2016, \mn@doi [\apj]
  {10.3847/0004-637X/820/1/12}, \href
  {http://adsabs.harvard.edu/abs/2016ApJ...820...12P} {820, 12}

\bibitem[\protect\citeauthoryear{{Prabhu} \& {Anupama}}{{Prabhu} \&
  {Anupama}}{2010}]{2010ASInC...1..193P}
{Prabhu} T.~P.,  {Anupama} G.~C.,  2010, in Astronomical Society of India
  Conference Series.

\bibitem[\protect\citeauthoryear{{Punsly}}{{Punsly}}{1999}]{1999ApJ...527..624P}
{Punsly} B.,  1999, \mn@doi [\apj] {10.1086/308091}, \href
  {http://adsabs.harvard.edu/abs/1999ApJ...527..624P} {527, 624}

\bibitem[\protect\citeauthoryear{{Rees}}{{Rees}}{1984}]{1984ARA&A..22..471R}
{Rees} M.~J.,  1984, \mn@doi [\araa] {10.1146/annurev.aa.22.090184.002351},
  \href {https://ui.adsabs.harvard.edu/abs/1984ARA&A..22..471R} {22, 471}

\bibitem[\protect\citeauthoryear{{Romero}, {Cellone}  \& {Combi}}{{Romero}
  et~al.}{1999}]{1999A&AS..135..477R}
{Romero} G.~E.,  {Cellone} S.~A.,   {Combi} J.~A.,  1999, \mn@doi [\aaps]
  {10.1051/aas:1999184}, \href
  {http://adsabs.harvard.edu/abs/1999A%26AS..135..477R} {135, 477}

\bibitem[\protect\citeauthoryear{{Romero}, {Cellone}, {Combi}  \& et
  al.}{{Romero} et~al.}{2002}]{2002A&A...390..431R}
{Romero} G.~E.,  {Cellone} S.~A.,  {Combi} J.~A.,   et al. 2002, \mn@doi [\aap]
  {10.1051/0004-6361:20020743}, \href
  {http://adsabs.harvard.edu/abs/2002A%26A...390..431R} {390, 431}

\bibitem[\protect\citeauthoryear{{Sagar}}{{Sagar}}{1999}]{1999CSci...77..643S}
{Sagar} R.,  1999, Current Science, \href
  {http://adsabs.harvard.edu/abs/1999CSci...77..643S} {77, 643}

\bibitem[\protect\citeauthoryear{{Sagar} et~al.,}{{Sagar}
  et~al.}{2011}]{2011CSci..101.1020S}
{Sagar} R.,  et~al., 2011, Current Science, \href
  {http://adsabs.harvard.edu/abs/2011CSci..101.1020S} {101, 1020}

\bibitem[\protect\citeauthoryear{{Sbarrato}, {Padovani}  \&
  {Ghisellini}}{{Sbarrato} et~al.}{2014}]{2014MNRAS.445...81S}
{Sbarrato} T.,  {Padovani} P.,   {Ghisellini} G.,  2014, \mn@doi [\mnras]
  {10.1093/mnras/stu1759}, \href
  {https://ui.adsabs.harvard.edu/abs/2014MNRAS.445...81S} {445, 81}

\bibitem[\protect\citeauthoryear{{Shankar}, {Dai}  \& {Sivakoff}}{{Shankar}
  et~al.}{2008}]{2008ApJ...687..859S}
{Shankar} F.,  {Dai} X.,   {Sivakoff} G.~R.,  2008, \mn@doi [\apj]
  {10.1086/591488}, \href
  {https://ui.adsabs.harvard.edu/abs/2008ApJ...687..859S} {687, 859}

\bibitem[\protect\citeauthoryear{{S{\k{a}}dowski} \&
  {Narayan}}{{S{\k{a}}dowski} \& {Narayan}}{2016}]{2016MNRAS.456.3929S}
{S{\k{a}}dowski} A.,  {Narayan} R.,  2016, \mn@doi [\mnras]
  {10.1093/mnras/stv2941}, \href
  {https://ui.adsabs.harvard.edu/abs/2016MNRAS.456.3929S} {456, 3929}

\bibitem[\protect\citeauthoryear{{Stalin}, {Gopal-Krishna}, {Sagar}  \& et
  al.}{{Stalin} et~al.}{2004}]{2004JApA...25....1S}
{Stalin} C.~S.,  {Gopal-Krishna} {Sagar} R.,   et al. 2004, \mn@doi [Journal of
  Astrophysics and Astronomy] {10.1007/BF02702287}, \href
  {https://ui.adsabs.harvard.edu/abs/2004JApA...25....1S} {25, 1}

\bibitem[\protect\citeauthoryear{{Stocke}, {Morris}, {Weymann}  \& et
  al.}{{Stocke} et~al.}{1992}]{1992ApJ...396..487S}
{Stocke} J.~T.,  {Morris} S.~L.,  {Weymann} R.~J.,   et al. 1992, \mn@doi
  [\apj] {10.1086/171735}, \href
  {https://ui.adsabs.harvard.edu/abs/1992ApJ...396..487S} {396, 487}

\bibitem[\protect\citeauthoryear{{Vestergaard}, {Wilkes}  \&
  {Barthel}}{{Vestergaard} et~al.}{2000}]{2000ApJ...538L.103V}
{Vestergaard} M.,  {Wilkes} B.~J.,   {Barthel} P.~D.,  2000, \mn@doi [\apjl]
  {10.1086/312805}, \href
  {https://ui.adsabs.harvard.edu/abs/2000ApJ...538L.103V} {538, L103}

\bibitem[\protect\citeauthoryear{{Wang}, {Jiang}, {Zhou}  \& et al.}{{Wang}
  et~al.}{2008}]{2008ApJ...676L..97W}
{Wang} J.,  {Jiang} P.,  {Zhou} H.,   et al. 2008, \mn@doi [\apjl]
  {10.1086/586893}, \href
  {https://ui.adsabs.harvard.edu/abs/2008ApJ...676L..97W} {676, L97}

\bibitem[\protect\citeauthoryear{{Welling}, {Miller}, {Brandt}  \& et
  al.}{{Welling} et~al.}{2014}]{2014MNRAS.440.2474W}
{Welling} C.~A.,  {Miller} B.~P.,  {Brandt} W.~N.,   et al. 2014, \mn@doi
  [\mnras] {10.1093/mnras/stu402}, \href
  {https://ui.adsabs.harvard.edu/abs/2014MNRAS.440.2474W} {440, 2474}

\bibitem[\protect\citeauthoryear{{Wykes} et~al.,}{{Wykes}
  et~al.}{2019}]{2019MNRAS.485..872W}
{Wykes} S.,  et~al., 2019, \mn@doi [\mnras] {10.1093/mnras/stz348}, \href
  {https://ui.adsabs.harvard.edu/abs/2019MNRAS.485..872W} {485, 872}

\bibitem[\protect\citeauthoryear{{Yang}, {Wu}, {Paragi}  \& {An}}{{Yang}
  et~al.}{2012}]{2012MNRAS.419L..74Y}
{Yang} J.,  {Wu} F.,  {Paragi} Z.,   {An} T.,  2012, \mn@doi [\mnras]
  {10.1111/j.1745-3933.2011.01182.x}, \href
  {http://adsabs.harvard.edu/abs/2012MNRAS.419L..74Y} {419, L74}

\bibitem[\protect\citeauthoryear{{Yuan} \& {Narayan}}{{Yuan} \&
  {Narayan}}{2014}]{2014ARA&A..52..529Y}
{Yuan} F.,  {Narayan} R.,  2014, \mn@doi [\araa]
  {10.1146/annurev-astro-082812-141003}, \href
  {https://ui.adsabs.harvard.edu/abs/2014ARA&A..52..529Y} {52, 529}

\bibitem[\protect\citeauthoryear{{Zhao}, {Chen}, {Shen}  \& et al.}{{Zhao}
  et~al.}{2015}]{2015AJ....149...46Z}
{Zhao} G.-Y.,  {Chen} Y.-J.,  {Shen} Z.-Q.,   et al. 2015, \mn@doi [\aj]
  {10.1088/0004-6256/149/2/46}, \href
  {https://ui.adsabs.harvard.edu/abs/2015AJ....149...46Z} {149, 46}

\bibitem[\protect\citeauthoryear{{Zhou}, {Wang}, {Wang}  \& et al.}{{Zhou}
  et~al.}{2006}]{2006ApJ...639..716Z}
{Zhou} H.,  {Wang} T.,  {Wang} H.,   et al. 2006, \mn@doi [\apj]
  {10.1086/499768}, \href {http://adsabs.harvard.edu/abs/2006ApJ...639..716Z}
  {639, 716}

\bibitem[\protect\citeauthoryear{{de Diego}}{{de
  Diego}}{2014}]{2014AJ....148...93D}
{de Diego} J.~A.,  2014, \mn@doi [\aj] {10.1088/0004-6256/148/5/93}, \href
  {http://adsabs.harvard.edu/abs/2014AJ....148...93D} {148, 93}

\makeatother
\end{thebibliography}
\label{lastpage}
\end{document}